\documentclass[preprint,12pt]{elsarticle}

\usepackage{amssymb}
\usepackage{amsmath}
\usepackage{amsfonts}

\usepackage{color}

\newtheorem{theorem}{Theorem}

\journal{}

\begin{document}

\begin{frontmatter}



\title{The Birkhoff theorem for unitary matrices of arbitrary dimensions}

\author[1,2]{Stijn De Baerdemacker}
\ead{stijn.debaerdemacker@ugent.be}
\author[3]{Alexis De Vos}
\ead{alexis.devos@elis.ugent.be}
\author[4,5]{Lin Chen}
\ead{linchen@buaa.edu.cn}
\author[7]{Li Yu}
\ead{yupapers@sina.com}
\address[1]{Ghent University, Center for Molecular Modeling, Technologiepark 903, 9052 Zwijnaarde, Belgium}
\address[2]{Ghent University, Ghent Quantum Chemistry Group, Krijgslaan 281, 9000 Gent, Belgium}
\address[3]{Ghent University, Cmst, Elis, Technologiepark 15, 9052 Zwijnaarde, Belgium}
\address[4]{School of Mathematics and Systems Science, Beihang University, Beijing 100191, China}
\address[5]{International Research Institute for Multidisciplinary Science, Beihang University, Beijing 100191, China}
\address[7]{National Institute of Informatics, 2-1-2 Hitotsubashi, Chiyoda-ku, Tokyo 101-8430, Japan}

\date{\today}

\begin{abstract}
It was shown recently that Birkhoff's theorem for doubly stochastic matrices can be extended to unitary matrices with equal line sums whenever the dimension of the matrices is prime.  We prove a generalization of the Birkhoff theorem for unitary matrices with equal line sums for arbitrary dimension.
\end{abstract}


\end{frontmatter}


\section{Introduction}\label{section:intro}
Let $M$ be an arbitrary $n \times n$ doubly-stochastic matrix.  Hence, each matrix entry $M_{jk}$ is a real number satisfying
$0 \le M_{jk} \le 1$ and each line sum, both $\sum_j M_{jk}$  and $\sum_k M_{jk}$ equal 1.  Then, the Birkhoff theorem \cite{birkhoff} tells us that $M$ can be written as a weighted sum of the $n!$ permutation matrices $P_m$ of dimension~$n$:
\begin{equation}
M = \sum_m c_m P_m
\end{equation}
such that all coefficients $c_m$ are real and satisfy $0 \le c_m \le 1$ and $\sum_m c_m~=~1$.

Let $U$ be an arbitrary $n \times n$ unitary matrix, such that each line sum equals~1, i.e.\  
$\sum_j U_{jk}=1$ and $\sum_k U_{jk}=1$.  Then, De Vos and De Baerdema\-cker \cite{unitarybirkhoff} conjectured that $U$ can be written as a weighted sum of the $n!$ permutation matrices $P_m$ of dimension~$n$:
\begin{equation}
U = \sum_m c_m P_m
\end{equation}
such that all coefficients $c_m$ are complex and satisfy both $\sum_m c_m = 1$ and $\sum_m |c_m|^2 = 1$.  They proved this fact for the case of prime~$n$.  In the present paper we demonstrate that the conjecture, which we will refer to as ``Conjecture 1'', is also valid for composite~$n$.

Before investigating the mathematics of the proof of Conjecture 1, we stress here three important differences between the `classical' Birkhoff theorem and the present `unitary' Birkhoff theorem:
\begin{itemize}
\item Whereas  the $n \times n$ doubly stochastic matrices form an $(n-1)^2$-dimensional semigroup, the $n \times n$ unit-linesum unitary matrices form an  $(n-1)^2$-dimensional group, isomorphic to the unitary group U($n-1$)
      and denoted XU($n$) \cite{xuzu}.
\item The unitary theorem is `stricter' than the classical theorem, in the sense that $\sum_m |c_m|^2 = 1$ implies $0 \le |c_m| \le 1$ for all~$m$, whereas  $0 \le |c_m| \le 1$ for all~$m$ does not necessarily imply $\sum_m |c_m|^2 = 1$.  
\item Let X($n$) be the $2(n-1)^2$-dimensional group of all invertible $n \times n$ matrices with complex entries and all line sums equal to~1.  The $(n-1)^2$-dimensional group XU($n$) is a compact subgroup of X($n$).  Also the $(n-1)^2$-dimensional semigroup DS($n$) of $n \times n$ doubly stochastic matrices resides within X($n$).  The intersection of the two subsets XU($n$) and DS($n$) consists of the finite (and thus 0-dimensional) group of $n \times n$ permutation matrices. Proof is in \ref{appendix:polytopes}.  Let $P$ and $Q$ be two arbitrary $n \times n$ permutation matrices.  The classical Birkhoff interpolation $(1-c)P + cQ$ with $c$ real and $0 \le c \le 1$ is a line segment in the $(n-1)^2$-dimensional space of DS($n$), constituting an edge of the so-called Birkhoff polytope \cite{bengtsson}.  In contrast, the quantum Birkhoff interpolation $(1-c)P + cQ$ with $c$ complex and $|c|^2 + |1-c|^2 = 1$  is a closed line in the compact $(n-1)^2$-dimensional space of XU($n$).  The line segment and the closed curve have only the points $P$ and $Q$ in common.  This also implies that the classical Birkhoff theorem is not a corollary of the unitary Birkhoff theorem.
\end{itemize}
The following sections are devoted to the proof and discussion of Conjecture 1 on the unitary Birkhoff theorem.  Section \ref{section:shur} recapitulates necessary basic ideas from representation theory, and sections \ref{section:birkhoff} \& \ref{section:explicit} present the proof of Conjecture 1.  Sections \ref{section:birkhoff} and \ref{section:explicit} are different in the sense that the former gives an existence proof, whereas the latter presents an explicit construction based on the proof.  We conclude the paper with a couple of examples in section \ref{section:examples}, and give our conclusions in section \ref{section:conclusion}.
\section{Schur decomposition of the permutation matrices}\label{section:shur}
The set of permutation matrices $\{P_\sigma|\sigma\in S_n\}$ forms a reducible representation of $S_n$.  It is straightforward to show that it decomposes into a direct sum of the trivial and standard representation, 
\begin{equation}
P = D^{(0)} \oplus D^{(s)},
\end{equation} 
by means of a similarity transformation with the $n$-dimensional Fourier matrix $F_n$.  The permutation matrices $P_\sigma$ form a discrete subgroup of XU$(n)$, and can therefore be brought into block diagonal form by means of \cite{negator}
\begin{equation}
F^\dag_n P_\sigma F_n=\left(\begin{array}{cc}1 & 0 \\ 0 & D^{(s)}(\sigma)\end{array}\right),\label{schur:FPF}
\end{equation}
with the trivial representation $D^{(0)}(\sigma)=1$ ($\forall\sigma\in S_n)$ on the upper-left block, and an $(n-1)$-dimensional representation $D^{(s)}$ on the lower-right block.  The latter representation can be identified with the so-called \emph{standard} (irreducible) representation for three reasons.  First, the representation is faithful, because the set of permutation matrices forms a faithful representation and $F_n$ is unitary.  Second, the representation is irreducible thanks to the orthogonality theorem on the characters of irreducible representations (see \ref{appendix:orthogonality})  \cite{hamermesh}.  Third, the characters of the irrep depend on the number of invariant points in the permutation ($\sigma(k)=k$), typical for the standard representation.
 As an example, we explicitly construct the standard representation of $S_3$ by means of the formula in eq. (\ref{schur:FPF}).  The permutation matrices $\{P_\sigma\}$ in $S_3$ are given by
\begin{align}
 &P_{0}=\left(\begin{array}{ccc} 1 & 0 & 0 \\ 0 & 1 & 0 \\ 0 & 0 & 1\end{array}\right),\  P_{(123)}=\left(\begin{array}{ccc} 0 & 0 & 1 \\ 1 & 0 & 0 \\ 0 & 1 & 0\end{array}\right),\  P_{(132)}=\left(\begin{array}{ccc} 0 & 1 & 0 \\ 0 & 0 & 1 \\ 1 & 0 & 0\end{array}\right),\\
 &P_{(12)}=\left(\begin{array}{ccc} 0 & 1 & 0 \\ 1 & 0 & 0 \\ 0 & 0 & 1\end{array}\right),\  P_{(13)}=\left(\begin{array}{ccc} 0 & 0 & 1 \\ 0 & 1 & 0 \\ 1 & 0 & 0\end{array}\right),\  P_{(23)}=\left(\begin{array}{ccc} 1 & 0 & 0 \\ 0 & 0 & 1 \\ 0 & 1 & 0\end{array}\right),\notag
\end{align}
and the Fourier transform $F_3$ is given by (with $\omega^3=1)$ 
\begin{equation}\label{schur:fourier}
F_3=\frac{1}{\sqrt{3}}\left(\begin{array}{ccc} 1 & 1 & 1 \\ 1 & \omega & \omega^2 \\ 1 & \omega^2 & \omega\end{array}\right),
\end{equation}
so the standard representation becomes
\begin{align}
 &D^{(s)}(0)=\left(\begin{array}{ccc} 1 & 0 \\ 0 & 1 \end{array}\right),\  D^{(s)}(123)=\left(\begin{array}{cc} \omega^2 & 0 \\ 0 & \omega \end{array}\right),\ D^{(s)}(132)=\left(\begin{array}{cc} \omega & 0 \\ 0 & \omega^2 \end{array}\right)\notag\\
 &D^{(s)}(12)=\left(\begin{array}{cc} 0 & \omega^2  \\ \omega & 0 \end{array}\right),\ D^{(s)}(13)=\left(\begin{array}{cc} 0 & \omega \\ \omega^2 & 0 \end{array}\right),\  D^{(s)}(23)=\left(\begin{array}{cc} 0 & 1 \\ 1 & 0 \end{array}\right).\label{schur:Ds3}
\end{align}
Note that the characters of the standard representation indeed correspond with the character table of $S_3$ (see table \ref{table:charactertables3}).
\begin{table}[!htb]
\begin{center}
\begin{tabular}{l|rrr}
 & $(1^3)$ & $(2,1)$& $(3)$\\
 \hline
$D^{(0)}$ & $1$ & $1$ & $1$ \\
$D^{(-)}$ & $1$ & $-1$ & $1$ \\
$D^{(s)}$ & $2$ & $0$ & $-1$
\end{tabular}
\caption{Character table of $S_3$ \cite{hamermesh}.  The conjugacy classes are labeled by their Young tableaux, and the irreps are the trivial irrep $D^{(0)}$, the sign irrep $D^{(-)}$ and the standard irrep $D^{(s)}$.  }\label{table:charactertables3}
\end{center}
\end{table}
\section{Birkhoff's theorem for unitary matrices}\label{section:birkhoff}
The existence proof of the unitary Birkhoff theorem is a corollary of two theorems, the first being provided by some of us \cite{unitarybirkhoff} and the second by Klappenecker \& R\"otteler \cite{klappenecker}.  
\begin{theorem}[De Vos \& De Baerdemacker \cite{unitarybirkhoff}]\label{theorem:dvdb}
If a matrix belongs to XU$(n)$, then it can be written as a weighted sum of
permutation matrices with the sum of the weights equal to 1.
\end{theorem}
This theorem was proven by induction in \cite{unitarybirkhoff}.  Before phrasing the second theorem, we need to introduce the concept of a \emph{group-circulant} matrix of a finite group $G$ of order $d$, associated to a vector $|c\rangle\in\mathbb{C}^d$.  Assume a one-to-one mapping between the elements $g\in G$ and the components $c_g$ of the vector $|c\rangle$, then the $(d\times d)$-matrix
\begin{equation}
\textrm{circ}_G(|c\rangle):=(c_{g^{-1}h})_{g,h\in G}
\end{equation}
is called the group-circulant for the group $G$, associated to $|c\rangle$.  The second theorem now states
\begin{theorem}[Klappenecker \& R\"otteler \cite{klappenecker}]\label{theorem:kr} Let $D$ be a matrix representation of a finite group $G$ ($g\rightarrow D(g)$).  If a unitary matrix $A$ can be expressed as linear combination
\begin{equation}
A=\sum_{g\in G}c_gD(g),\qquad c_g\in\mathbb{C},
\end{equation}
then the coefficients $c_g$ of the vector $|c\rangle$ can be chosen such that the associated group-circulant matrix is also unitary.
\end{theorem}
These two theorems essentially prove the Birkhoff theorem for unitary matrices of arbitrary dimensions.  Indeed, the permutation matrices form a matrix representation of the symmetric group for any dimension.  Theorem \ref{theorem:dvdb} ensures that any matrix $X\in\textrm{XU}(n)$ can be expressed as a linear combination of the permutation matrices
\begin{equation}
X=\sum_{\sigma\in S_n}c_\sigma P_\sigma,
\end{equation}
so the two prerequisites of Theorem \ref{theorem:kr} are met.  Therefore, one can choose the vector $|c\rangle$ such that its associated group circulant matrix $\textrm{circ}_{S_n}(|c\rangle)$ is unitary.    More specific, this implies that $|c\rangle$ can be chosen such that
\begin{equation}
\sum_{\sigma\in S_n}|c_\sigma|^2=1.
\end{equation}
This essentially proves our Conjecture 1 on the Birkhoff theorem for unitary matrices.  However, we will go one step further and present explicit expressions for the $c_\sigma$ coefficients in Theorem \ref{theorem:explicit} in the following section.

Before presenting Theorem \ref{theorem:explicit}, it is useful to recapitulate the key concepts in the proof of Theorem  \ref{theorem:kr} by Klappenecker \& R\"otteler \cite{klappenecker}.  The proof relies entirely on the Schur orthonormality of the irreps of finite groups.  Just like in the previous paper \cite{unitarybirkhoff}, it is most instructive to consider the $n=3$ example before illustrating the general case.  The claim is that any $X\in\textrm{XU}(3)$ can be written as
\begin{equation}\label{birkhoff:X3}
X=\sum_{\sigma\in S_3}c_\sigma P_\sigma.
\end{equation} 
with $\sum_{\sigma\in S_3}c_\sigma=1$ and $\sum_{\sigma\in S_3}|c_{\sigma}|^2=1$.  Performing a similarity transformation with the $F_3$ Fourier matrix of eq.\ (\ref{schur:fourier}) on the lhs and rhs of eq.\ (\ref{birkhoff:X3}) gives rise to
\begin{equation}\label{birkhoff:FX3F}
\left(\begin{array}{cc}1 & 0 \\
                        0 & U^{(s)} \end{array}\right)=\sum_{\sigma\in S_3}c_\sigma \left(\begin{array}{cc}1 & 0 \\
                        0 & D^{(s)}(\sigma) \end{array}\right),
\end{equation}
with $U^{(s)}\in U(2)$, and the standard representation $D^{(s)}(\sigma)$ of $S_3$ given by the matrices (\ref{schur:Ds3}).  By equating all matrix elements explicitly, this gives rise to the following set of (five) equations
\begin{align}
1&=c_0+c_{(123)}+c_{(132)}+c_{(12)}+c_{(13)}+c_{(23)},\label{birkhoff:FX3Feq1}\\
U^{(s)}_{11}&=c_0+\omega^2 c_{(123)}+\omega c_{(132)},\\
U^{(s)}_{22}&=c_0+\omega c_{(123)}+\omega^2 c_{(132)},\\
U^{(s)}_{12}&=c_{(23)}+ \omega^2 c_{(12)}+\omega c_{(13)}, \\
U^{(s)}_{21}&=c_{(23)}+ \omega c_{(12)}+\omega^2 c_{(13)}\label{birkhoff:FX3Feq5}.
\end{align}
Note that the first equation $1=\sum_{\sigma}c_{\sigma}$ confirms Theorem \ref{theorem:dvdb}.  The linear set of equations contains six variables, so one needs one additional equation to find a closed solution.  Remark that the sign representation $D^{(-)}(\sigma)$ is the only irrep missing in eq.\ (\ref{birkhoff:FX3F}).  Indeed, one can augment the $3\times 3$ matrix equality to a $4\times 4$ matrix equality by adding the sign representation $D^{(-)}(\sigma)$ on the diagonal
\begin{equation}\label{birkhoff:FX3F+}
\left(\begin{array}{ccc}1 & 0 & 0\\
                        0 & U^{(s)} & 0 \\
                        0 & 0 & e^{i\phi} \end{array}\right)=\sum_{\sigma\in S_3}c_\sigma \left(\begin{array}{ccc}D^{(0)}(\sigma) & 0 & 0 \\
                        0 & D^{(s)}(\sigma) & 0 \\
                        0 & 0 & D^{(-)}(\sigma) \end{array}\right),
\end{equation}
where we have used that $1=D^{(0)}(\sigma)$ for all $\sigma$, and $e^{i\phi}\in U(1)$ is an arbitrary $1\times 1$ unitary matrix.  This extra equation reads
\begin{equation}
e^{i\phi}=c_0+c_{(123)}+c_{(132)}-c_{(12)}-c_{(13)}-c_{(23)}\label{birkhoff:FX3Feq6}.
\end{equation}
Solving eqs.\ (\ref{birkhoff:FX3Feq1}-\ref{birkhoff:FX3Feq5},\ref{birkhoff:FX3Feq6}) for $c_\sigma$, we obtain
\begin{align}
c_0&=\tfrac{1}{3}[\tfrac{1}{2}(1+e^{i\phi})+U^{(s)}_{11}+U^{(s)}_{22}]\\
c_{(123)}&=\tfrac{1}{3}[\tfrac{1}{2}(1+e^{i\phi})+\omega U^{(s)}_{11}+\omega^2U^{(s)}_{22}]\\
c_{(132)}&=\tfrac{1}{3}[\tfrac{1}{2}(1+e^{i\phi})+\omega^2 U^{(s)}_{11}+\omega U^{(s)}_{22}]\\
c_{(23)}&=\tfrac{1}{3}[\tfrac{1}{2}(1-e^{i\phi})+U^{(s)}_{12}+U^{(s)}_{21}]\\
c_{(12)}&=\tfrac{1}{3}[\tfrac{1}{2}(1-e^{i\phi})+\omega U^{(s)}_{12}+\omega^2U^{(s)}_{21}]\\
c_{(13)}&=\tfrac{1}{3}[\tfrac{1}{2}(1-e^{i\phi})+\omega^2 U^{(s)}_{12}+\omega U^{(s)}_{21}].
\end{align}
Taking into account that $U^{(s)}$ is unitary, it is straightforward to verify that $\sum_\sigma |c_\sigma|^2=1$, which again proves the unitary Birkhoff theorem for $n=3$.  Note that the choice $\phi=0$ coincides exactly with the choice $p=1$ in \cite{unitarybirkhoff}.  More generally, the correspondence between the present solution and the one presented in \cite{unitarybirkhoff}, is $p=\frac{1}{2}(1+e^{i\phi})$, coinciding with the observation that $p$ had to lie on a circle in the complex plane with radius $\frac{1}{2}$ and center $\frac{1}{2}$.  

For the proof of the general case, we proceed in a similar manner.  Theorem \ref{theorem:dvdb} states that any $X\in\textrm{XU}(n)$ can be written as
\begin{equation}
X=\sum_{\sigma\in S_n}c_{\sigma}P_\sigma.
\end{equation} 
Applying the similarity transform with the $F_n$ Fourier transform, we obtain
\begin{equation}
\left(\begin{array}{cc}1 & 0 \\
                        0 & U^{(s)} \end{array}\right)=\sum_{\sigma\in S_n}c_\sigma \left(\begin{array}{cc}D^{(0)}(\sigma) & 0 \\
                        0 & D^{(s)}(\sigma) \end{array}\right),
\end{equation}
with $U^{(s)}\in\textrm{U}(n-1)$.  This results in a set of $(n-1)^2+1$ equations with $n!$ variables $c_\sigma$, so we need $n!-(n-1)^2-1$ more equations to find a closed solution via linear algebra.  Now, one can augment the $n\times n$ matrix with all other possible irreps $D^{(\nu)}$ of $S_n$ on the (block) diagonal in the rhs summation, and arbitrary $n_\nu$ dimensional unitary matrices $U^{(\nu)}\in \textrm{U}(n_\nu)$ on the corresponding (block) diagonal parts in the lhs, with $n_\nu$ the dimension of irrep $D^{(\nu)}$
\begin{equation}
\left(\begin{array}{cccc}1 &  & & \\
                         & U^{(s)} & & \\
                         &  & U^{(2)} & \\
                         &  &    & \ddots \end{array}\right)=\sum_{\sigma\in S_n}c_\sigma \left(\begin{array} {cccc}D^{(0)}(\sigma) &  &  & \\
                         & D^{(s)}(\sigma) &  & \\
                         &  & D^{(2)}(\sigma) & \\
                         &  &  &  \ddots \end{array}\right).\label{birkhoff:regularrep}
\end{equation}
The resulting matrix has dimension $d(n)=\sum_{\nu=0}^{k-1}n_\nu$ with $k$ the number of conjugacy classes of $S_n$, i.e., the number of partitions of integer $n$ \cite{integersequence:a000041}.  The number $d(n)$ equals the number of involutions on $n$ objects and is tabulated in \cite{integersequence:a000085}.  The augmented matrices in the rhs summation in (\ref{birkhoff:regularrep}) form again a reducible representation, known as the \emph{regular} representation \cite{burrow}.  We will label the standard representation $D^{(s)}=D^{(1)}$ from now on for the sake of notation.  

The key observation is that there are as many irrep matrix elements $D^{(\nu)}_{ij}(\tau)$ ($\nu=0,\dots, k-1$ \& $i,j=1,\dots n_\nu$) for any $\tau\in S_n$ as there are group elements $\sigma$ in $S_n$.  Therefore, $D^{(\nu)}_{ij}(\sigma)$ can be regarded as a matrix element of an $n!\times n!$ matrix $D$ with indices $\{\nu,i,j\}$ for the ``rows'' and $\sigma$ for the ``columns''.  Due to Schur's orthogonality theorems, 
\begin{align}
n_\nu\sum_{\sigma\in S_n} D^{(\nu)\ast}_{ij}(\sigma) D^{(\mu)}_{lm}(\sigma)&=n!\delta_{\mu\nu}\delta_{il}\delta_{jm}\label{birkhoff:schurortho1}\\
\sum_{\nu=0}^{k-1}n_\nu\sum_{ij=1}^{n_\nu}D^{(\nu)\ast}_{ij}(\sigma)D^{(\nu)}_{ij}(\tau)&=n!\delta_{\sigma\tau}\label{birkhoff:schurortho2}
\end{align}
it follows directly that the (rescaled) matrix $\tilde{D}$ is a unitary matrix, with $\tilde{D}_{ij}^{(\nu)}(\sigma)=\sqrt{\frac{n_\nu}{n!}}D_{ij}^{(\nu)}(\sigma)$.  This means that the set of $n!$ equations
\begin{equation}
U^{(\nu)}_{ij}=\sum_{\sigma\in S_n}c_\sigma D^{(\nu)}_{ij}(\sigma),\quad \forall \nu,i,j,
\end{equation}
extracted from the block diagonals of eq. (\ref{birkhoff:regularrep}) can be solved uniquely by inverting the non-singular matrix $D$.  This is exactly what has been done in Theorem \ref{theorem:explicit} in the following section to construct explicit expressions for the coefficients $c_\sigma$ in the Birkhoff summation.
\section{Explicit decomposition}\label{section:explicit}
\begin{theorem}\label{theorem:explicit}
Every matrix $X\in XU(n)$ can be written as a weighted sum of permutation matrices
\begin{equation}
X=\sum_{\sigma\in S_n} c_\sigma P_\sigma,\label{birkhoff:theorem3:decomposition}
\end{equation}
with the coefficients $c_\sigma$ given by
\begin{align}
c_\sigma&=\frac{1}{n!}\sum_{\nu}n_\nu\sum_{ij=1}^{n_\nu}D_{ij}^{(\nu)\ast}(\sigma)U_{ij}^{(\nu)},\label{birkhoff:theorem3:c}\\
&=\frac{1}{n!}\sum_{\nu}n_\nu\emph{Tr}(D^{(\nu)\dag}(\sigma)U^{(\nu)})\label{birkhoff:theorem3:ctrace}
\end{align}
where the sum $\nu$ runs over all possible irreps $D^{(\nu)}$ of $S_n$, $n_\nu$ is the dimension of $D^{(\nu)}$, and $U^{(\nu)}\in U(n_\nu)$ is a unitary matrix associated to $D^{(\nu)}$.  All unitary matrices $U^{(\nu)}$ can be chosen arbitrarily, with the exception of $U^{(0)}=1\in U(1)$ and $U^{(1)}=U^{(s)}\in U(n-1)$ associated to respectively the trivial and standard representation, which are found by 
\begin{equation}
F^\dag_n X F_n =\left(\begin{array}{cc} U^{(0)} & 0 \\ 0 & U^{(s)} \end{array}\right).\label{birkhoff:theorem3:FXF}
\end{equation}
In addition, the sum of the moduli squared equals 1
\begin{equation}
\sum_{\sigma\in S_n}|c_\sigma|^2 =1.\label{birkhoff:theorem2:c2}
\end{equation} 
\end{theorem}
The proof goes by straightforward application of Shur's orthogonality relations.  First, we verify that eq.\ (\ref{birkhoff:theorem3:FXF}) holds when the $c_\sigma$ coefficients are taken as in expression (\ref{birkhoff:theorem3:ctrace}) (from which $X$ follows directly by means of the similarity transform $F_n(F^\dag_nXF_n)F^\dag_n=X$.).  Because of the Schur decomposition (\ref{schur:FPF}) of the permutation matrices $P_\sigma$, $F^\dag_n X F_n$ is also block diagonal.  We first investigate the upper left matrix element
\begin{equation}
(F^\dag_n X F_n )_{11}=\sum_{\sigma\in S_n}c_\sigma(F^\dag_n P_\sigma F_n )_{11}=\sum_{\sigma\in S_n}c_\sigma D^{(0)}_{11}(\sigma)\label{birkhoff:theorem3:upperleft}
\end{equation}
in which we prefer to keep the explicit notation for the trivial representation $D^{(0)}_{11}(\sigma)=1$ in the Schur decomposition of the permutation matrices (\ref{schur:FPF}).  We now insert the explicit expression (\ref{birkhoff:theorem3:ctrace}) for $c_\sigma$ in (\ref{birkhoff:theorem3:upperleft}) and rearrange the summation
\begin{equation}
(F^\dag_n X F_n )_{11}=\frac{1}{n!}\sum_{\nu}n_\nu\sum_{ij=1}^{n_\nu}U_{ij}^{(\nu)}\sum_{\sigma\in S_n}D_{ij}^{(\nu)\ast}(\sigma)D^{(0)}_{11}(\sigma)=U^{(0)}_{11},
\end{equation}
where the Schur orthogonality relation (\ref{birkhoff:schurortho1}) has been used in the last step.  Because $U^{(0)}\equiv1$ in the present theorem, we get that the upper left element of $F^\dag_n X F_n$ equals 1.  The other matrix elements can be obtained analogously from Schur's orthogonality relation (\ref{birkhoff:schurortho1})
\begin{equation}
(F^\dag_n X F_n )_{lm}=U^{(1)}_{l-1,m-1}\equiv U^{(s)}_{l-1,m-1},\qquad \forall l,m=2,\dots, n.
\end{equation}

The compactness of the $c_\sigma$ coefficients (\ref{birkhoff:theorem2:c2}) can also be proven similarly
\begin{align}
\sum_{\sigma\in S_n}|c_\sigma|^2&=\sum_{\sigma\in S_n}c_\sigma c_\sigma^\ast\\
&=\frac{1}{n!^2}\sum_{\nu\mu}n_\nu n_\mu \sum_{ij=1}^{n_\nu}\sum_{lm=1}^{n_\mu}U_{ij}^{(\nu)}U_{lm}^{(\mu)\ast}\sum_{\sigma\in S_n}D_{ij}^{(\nu)\ast}(\sigma)D_{lm}^{(\mu)}(\sigma)\\
&=\frac{1}{n!}\sum_{\nu}n_\nu \sum_{ij=1}^{n_\nu}U_{ij}^{(\nu)}U_{ij}^{(\nu)\ast}\\
&=\frac{1}{n!}\sum_{\nu}n_\nu \textrm{Tr}(U^{(\nu)}U^{(\nu)\dag}).
\end{align}
Because $U^{(\nu)}$ are unitary, we have that $\textrm{Tr}(U^{(\nu)}U^{(\nu)\dag})=n_\nu$, so
\begin{equation}
\sum_{\sigma\in S_n}|c_\sigma|^2=\frac{1}{n!}\sum_{\nu}n_\nu^2=1.
\end{equation}
This completes the proof.

Theorem \ref{theorem:explicit} allows for a large freedom of choice, associated with all unitary matrices $U^{(\nu)}$ ($\nu>1$) that can be chosen arbitrarily.  As an example, we investigate the choice $U^{(\nu)}$ (for $\nu >1$) equal to $D^{(\nu)}(\tau)$, where $\tau$ is one particular permutation, i.e.\ a particular member of $S_n$. Eq.\ (\ref{birkhoff:theorem3:ctrace}) thus becomes
\begin{align}
c_{\sigma} 
& =\frac{1}{n!}\ \sum_{\nu=0}^{k-1} n_{\nu} \mbox{Tr} \left(D^{(\nu)}(\sigma)^\dag D^{(\nu)}(\tau)\right)\notag\\
&\quad- \frac{n_1}{n!}\mbox{Tr} \left(D^{(1)}(\sigma)^\dag D^{(1)}  (\tau)\right)+ \frac{n_1}{n!} \mbox{Tr} \left(D^{(1)}  (\sigma)^\dag U^{(1)}        \right).
\end{align}
Taking into account that $n_1=n-1$, $D^{(1)}(\sigma)^\dag=D^{(1)}(\sigma^{-1})$, and that the first term in the equation is Shur's orthogonality relation, we obtain
\begin{equation}
c_{\sigma} = \delta_{\sigma\tau} - \frac{n-1}{n!}\ \chi^{(1)}(\sigma^{-1} \tau)
             + \frac{n-1}{n!}\ \mbox{Tr} \left( D^{(1)}(\sigma^{-1}) U^{(1)} \right).\label{explicit:cstandardirrep}
\end{equation}
For instance, substituting $n=3$ and $\tau=(0)$ the identity, we immediately recover (19-24) with $\phi=0$.

Note that in contrast to eq.\ (\ref{birkhoff:theorem3:ctrace}), only the standard representation $D^{(s)}$ occurs in the expression (\ref{explicit:cstandardirrep}) for the $c_\sigma$ coefficient.  From a practical point of view, this is very convenient because the standard representation can quickly be obtained by reducing the $n$-dimensional permutation representation (\ref{schur:FPF}).  Hence, no other representations need to be constructed from other means.

If $n>3$, then another elegant choice is possible.  Again, we choose $U^{(\nu)}=D^{(\nu)}(\tau)$, however, now with \emph{two} exceptions: both $U^{(s)}=U$ and $U^{(a)}=U$. Here, $U^{(s)}$ corresponds, in (\ref{birkhoff:regularrep}), with the standard representation $D^{(s)}$ of dimension $n-1$, whereas $U^{(a)}$ corresponds with the other $(n-1)$-dimensional representation $D^{(a)}$ in (\ref{birkhoff:regularrep}), which we will call the `anti-standard' representation.  The anti-standard representation of $S_n$ consists of the same matrices as the standard representation, except for a minus sign in case of an odd permutation\footnote{Also for $n=2$ and $n=3$ an anti-standard representation exists. However, for $n=2$, the anti-standard representation equals the trivial representation and for $n=3$,  the anti-standard representation is equivalent to the standard representation:
                     \[
                     D^{(a)}(\sigma) = 
                     {\tiny \left( \begin{array}{cc} 1 & \\ & -1 \end{array} \right)}
                     \ D^{(s)}(\sigma)\ 
                     {\tiny \left( \begin{array}{cc} 1 & \\ & -1 \end{array} \right)}.
                     \]
                    }.
Now, eq.\ (\ref{birkhoff:theorem3:ctrace}) becomes
\begin{align}
c_{\sigma} 
 = \delta_{\sigma\tau}&- \frac{n-1}{n!}\chi^{(s)}(\sigma^{-1}\tau)
+ \frac{n-1}{n!}\mbox{Tr} \left(D^{(s)}  (\sigma^{-1}) U^{(1)}        \right)\notag \\
&  
- \frac{n-1}{n!}\chi^{(a)}(\sigma^{-1}\tau)+ \frac{n-1}{n!}\mbox{Tr} \left(D^{(a)}  (\sigma^{-1}) U^{(1)}        \right). 
\end{align}
Restricting ourselves to the obvious choice $\tau=(0)$, we obtain,
\begin{equation}
c_{\sigma} = \left\{\begin{array}{ll}\delta_{\sigma 0} 
             - 2 \ \frac{n-1}{n!}\ \chi^{(s)}(\sigma)
             + 2 \ \frac{n-1}{n!}\ \mbox{Tr} \left( D^{(s)}(\sigma^{-1}) U \right), & \sigma\ \mbox{even}\\
             0,&\sigma\ \mbox{odd.}\end{array}\right.\label{birkhoff:decomposition:even}
\end{equation}

Yet another choice is $U^{(s)}=U$ and $U^{(a)}=-U$. It yields
\begin{equation}
c_{\sigma} = \delta_{\sigma 0} 
             - 2 \ \frac{n-1}{n!}\ \chi^{(s)}(\sigma)
\end{equation}
for even $\sigma$ and
\begin{equation}
c_{\sigma} =   2 \ \frac{n-1}{n!}\ \mbox{Tr} \left( D^{(s)}(\sigma^{-1}) U \right)
\end{equation}
for odd $\sigma$.  Again, only the standard representation $D^{(s)}$ is required.
\section{Examples}\label{section:examples}
We now present two examples for the non-prime $n=4$ case.  In the framework of quantum multiports \cite{mattle}, it is important to synthesize $n \times n$ matrices with all entries having the same modulus (and thus modulus equal to $1/\sqrt{n}$).  As an example, we consider the XU(4) matrix
\begin{equation}
X = 
\frac{1}{2}\ \left( \begin{array}{rrrr}
 1 & -i &  1 &  i \\
 1 &  1 & -1 &  1 \\
 1 &  i &  1 & -i \\
-1 &  1 &  1 &  1    \end{array} \right) \ .
\end{equation}
Up to phase changes, it equals the $4 \times 4$ Fourier transform.  Indeed:
\begin{equation}
\left( \begin{array}{rrrr}
 1 &   &   &    \\
   & 1 &   &    \\
   &   & 1 &    \\
   &   &   & -1 \end{array} \right) \ X \ 
\left( \begin{array}{rrrr}
 1 &   &   &    \\
   & i &   &    \\
   &   & 1 &    \\
   &   &   & -i \end{array} \right) = F_4\ .
\end{equation}

Applying (\ref{birkhoff:decomposition:even}) to $X$ yields, besides the twelve zero coefficients for the odd permutation matrices, the following weights for the twelve even permutation matrices:
\begin{equation}
\frac{1}{8}\ \{4, -1-i, 1+i, 2-2i, -1-i, -1-i, 1+i, -1-i, 4, 1+i, 1+i, -2+2i\} \ ,
\end{equation}
satisfying $\sum_{\sigma}\, c_{\sigma} = \sum_{\sigma}\, |c_{\sigma}|^2 = 1$.

Because a $4 \times 4$ permutation matrix is also an XU(4) matrix, we can e.g.\ apply (\ref{birkhoff:decomposition:even}) to the odd permutation matrix
\begin{equation}
 \left( \begin{array}{cccc}
1 & 0 & 0 & 0 \\
0 & 1 & 0 & 0 \\
0 & 0 & 0 & 1 \\
0 & 0 & 1 & 0 \end{array} \right) \ ,
\end{equation}
resulting in a Birkhoff decomposition with even permutation matrices with weights
\begin{equation}
\frac{1}{4}\ \{2, 1, 1, 2, -1, -1, -1, 1, 0, -1, 1, 0\} \ .
\end{equation}
Again we have $\sum_{\sigma}\, c_{\sigma} = \sum_{\sigma}\, |c_{\sigma}|^2 = 1$.  It is remarkable that, for $n>3$, one can decompose any permutation, either even or odd, into a sum of only even permutation matrices.
\section{Conclusion}\label{section:conclusion}
Recently, we proved a Birkhoff theorem for unitary matrices of prime dimension \cite{unitarybirkhoff}.  In the present manuscript, we complete the proof of the theorem for unitary matrices of arbitrary dimension $n$.  The proof is based on a theorem by Klappenecker and R\"otteler \cite{klappenecker}, employing the Schur orthonormality of the representations of the permutation group $S_n$.  Furthermore, we present an explicit construction of the weight coefficients in the Birkhoff decomposition.  There is a freedom of choice in the explicit construction, allowing one to express the coefficients exclusively in terms of the known $(n-1)$-dimensional standard representation.  Remarkably, this leads to a decomposition of any unitary matrix, for $n>3$, into a weighted sum of only even permutation matrices.  It would be interesting to explore the freedom of choice inherent in the procedure in order to obtain different and possibly shorter Birkhoff decompositions. 
\section*{Acknowledgements}
LC was supported by the NSF of China (Grant No.\ 11501024), and the Fundamental Research Funds for the Central Universities (Grant Nos.\ 30426401 and 30458601). LY was supported by NICT-A (Japan).
\appendix

\section{Unitary doubly stochastic matrices}\label{appendix:polytopes}

\begin{theorem}
The only unitary doubly stochastic matrices are the permutation matrices.
\end{theorem}
Proof.  Let $M$ be an $n \times n$ matrix that is simultaneously member of XU($n$) and DS($n$). Because $M \in$~DS($n$), all its entries are either zero or positive.  Every row of $M$ contains at least one non-zero entry (because otherwise the related row sum would equal~0).  The row cannot contain more than one non-zero entry.  Indeed, suppose the $a$th row contains two positive entries 
$M_{ab}$ and $M_{ac}$ ($b\neq c$). Then the sum $\sum_j M_{jb}M_{jc}$ contains at least one positive term, i.e.\ $M_{ab}M_{ac}$. All the other $n-1$ terms in the summation are either zero or positive, so we have $\sum_j M_{jb}M_{jc} > 0$.  

Because $M$ is a unitary matrix, we also have that $\sum_{j} (M^\dag)_{kj} M_{jl}=\delta_{kl}$, or $\sum_{j} M_{jb} M_{jc}=0$, which contradicts the previous paragraph.

Hence, every row of $M$ contains exactly one non-zero entry, and so does every column.  Because all line sums are equal to~1, these non-zero entries necessarily equal~1.  We conclude that $M$ is a permutation matrix.

\section{Reducibility of the permutation matrices}\label{appendix:orthogonality}
The orthogonality theorem on the characters of irreducible representations (see section 3-16. in \cite{hamermesh}) states that a representation $D^{(\nu)}$ is irreducible iff the sum of all characters squared $|\chi^\nu(\sigma)|^2$ over all possible group elements equals the order of the group.  For the symmetric group $S_n$, this becomes
\begin{equation}\label{appendixa:characterformula}
\sum_{\sigma\in S_n}|\chi^{(\nu)}(\sigma)|^2=n!. 
\end{equation} 
So, if this relation holds for the representation $D^{(s)}$ in eq.\ (\ref{schur:FPF}), it is irreducible.  The character 
\begin{equation}
\chi^{(s)}(\sigma)=\textrm{Tr}[D^{(s)}(\sigma)]
\end{equation}
can be obtained for each permutation $\sigma$ explicitly.  We indeed have that
\begin{equation}
\mbox{Tr}\, [D^{(s)}(\sigma)] = \mbox{Tr}\, ( F^{\dagger} P_{\sigma} F ) - 1,
\end{equation}
because $D^{(s)}(\sigma)$ is obtained from $F^{\dagger} P_{\sigma} F$
by deleting its first row and first column.
Because a trace is similarity-invariant, this yields
\begin{equation}
\chi^{(s)}(\sigma) = \mbox{Tr}\, (P_{\sigma}) - 1.
\end{equation}
As a result, the characters depend on the number of ones on the diagonal in the original matrix $P_{\sigma}$ only, in a way typical for the standard representation.  Thus, the characters $\chi^{(s)}$ are the rational integers ranging from $-1$ to $n-1$, except $n-2$.  The character formula (\ref{appendixa:characterformula}) becomes
\begin{align}
\sum_{\sigma\in S_n}|\chi^{(s)}(\sigma)|^2=\sum_{\sigma\in S_n}\left(\sum_{l=1}^n\delta_{l\sigma(l)}-1\right)\left(\sum_{k=1}^n\delta_{k\sigma(k)}-1\right)\\
=\sum_{l=1}^n\sum_{k=1}^n\sum_{\sigma\in S_n}\delta_{l\sigma(l)}\delta_{k\sigma(k)}-2\sum_{k=1}^n\sum_{\sigma\in S_n}\delta_{k\sigma(k)}+n!.
\end{align}
Taking into account that $\delta_{k\sigma(k)}^2=\delta_{k\sigma(k)}$, this can be slightly rewritten as
\begin{equation}
\sum_{\sigma\in S_n}|\chi^{(s)}(\sigma)|^2=\sum_{l=1}^n\sum_{k\neq l}^n\sum_{\sigma\in S_n}\delta_{l\sigma(l)}\delta_{k\sigma(k)}-\sum_{k=1}^n\sum_{\sigma\in S_n}\delta_{k\sigma(k)}+n!.
\end{equation}
Both sums are quite straightforward to reason.  There are exactly $(n-1)!$ permutations $\sigma$ in $S_n$ that leave $k$ invariant $\sigma(k)=k$, leading to
\begin{equation}
\sum_{k=1}^n\sum_{\sigma\in S_n}\delta_{k\sigma(k)}=\sum_{k=1}^n(n-1)! = n!.
\end{equation}
Similarly, there are exactly $(n-2)!$ permutations $\sigma$ in $S_n$ that leave $k$ and $l(\neq k)$ invariant, leading to
\begin{equation}
\sum_{l=1}^n\sum_{k\neq l}^n\sum_{\sigma\in S_n}\delta_{l\sigma(l)}\delta_{k\sigma(k)}=\sum_{l=1}^n\sum_{k\neq l}^n(n-2)!=n!.
\end{equation}
As a result, the character orthogonality formula (\ref{appendixa:characterformula}) becomes
\begin{equation}
\sum_{\sigma\in S_n}|\chi^{(s)}(\sigma)|^2=n!-n!+n!\equiv n!,
\end{equation}
pointing out that the representation $D^{(s)}$ is indeed irreducible.



\end{document}